\documentclass[
 reprint,
 amsmath,amssymb,
 aps,
]{revtex4-1}
\usepackage[normalem]{ulem}
\usepackage[italic]{hepnames}
\usepackage{natbib}
\usepackage{graphicx}
\usepackage{dcolumn}
\usepackage{bm}
\usepackage{amsmath}
\usepackage{hyperref}

\begin{document}

\preprint{APS/123-QED}

\title{Sterile neutrino search in the NOvA Far Detector}

\author{Sijith Edayath}
\email{sedayath@fnal.gov}
\affiliation{Cochin University of Science and Technology, India.}%

\author{Prof.\ Adam Aurisano, Prof.\ Alexandre Sousa , Shaokai Yang}
\affiliation{
 University of Cincinnatti, Cincinnatti, USA.\\
}

\author{Dr.\ Gavin S Davies}
\affiliation{
 Indiana University, USA.\\
}

\author{Dr.\ Louise Suter}
\affiliation{
 Fermi National Accelerator Laboratory, USA.\\
}

\collaboration{For the NOvA Collaboration}
\begin{abstract}
This document describes the search for sterile neutrinos in the NOvA Far Detector
through the Neutral Current disappearance channel. It also explains the improvements being readied for the 2017 analysis.

\end{abstract}

\maketitle

\section{Introduction}

The majority of neutrino oscillation experiments have obtained evidence for
neutrino oscillations that are compatible with the three-flavor model.
Explaining anomalous results from short-baseline experiments, such as LSND and MiniBooNE,
in terms of neutrino oscillations requires the existence of sterile neutrinos. 
The search for sterile neutrino mixing conducted in NOvA uses the Near Detector (ND)
at Fermilab and Far Detector (FD) in Minnesota.
The signal for sterile neutrino oscillations is a deficit of neutral-current neutrino
interactions at the FD with respect to the ND prediction. 
In this document, a summary of the 2016 analysis \cite{nc_paper} result and the analysis improvements that we are implementing for future NC sterile neutrino searches
with NOvA are summarized. These include: improved modeling of our detector response;  the inclusion of NC 2p2h interaction modeling;
implementing better energy reconstruction techniques; and including possible oscillation due to sterile 
neutrinos in the ND. These improvements enable us to do a simultaneous ND-FD shape fit of the NC energy
spectrum covering a wider
sterile mass range than previous analyses. 
\section{\label{sec:level1}Sterile neutrinos}
\subsubsection{LSND Anomaly}
Short-BaseLine (SBL) experiments such as LSND\cite{lsnd} and MicroBooNE\cite{microboone1, microboone2} measured an excess of $\APnue$ in $\APnum$ beam in the vicinity of $\frac{1 km}{GeV}$. This can be interpreted as a result of oscillation at a mass squared splitting 1~$eV^{2}$. It is not consistent with the 3-flavor mass splittings $\Delta m_{21}^{2}$ and $\Delta m_{32}^{2}$. The presence of three $\Delta m^{2}$ requires the existence of at least four neutrinos.
\subsubsection{Sterile Neutrino}
 LEP measurements of the $Z^{0}$ boson decay into
neutrinos is consistent with only 3 active neutrino flavors. If a fourth neutrino exists, it is either heavier than $\frac{M_{z^{0}}}{2}$, or otherwise it does not participate in the weak interaction; a Light Sterile Neutrino.
\subsubsection{Tension in the sterile neutrino searches}
\begin{figure}[tbp]
\centering
\includegraphics[width=\linewidth]{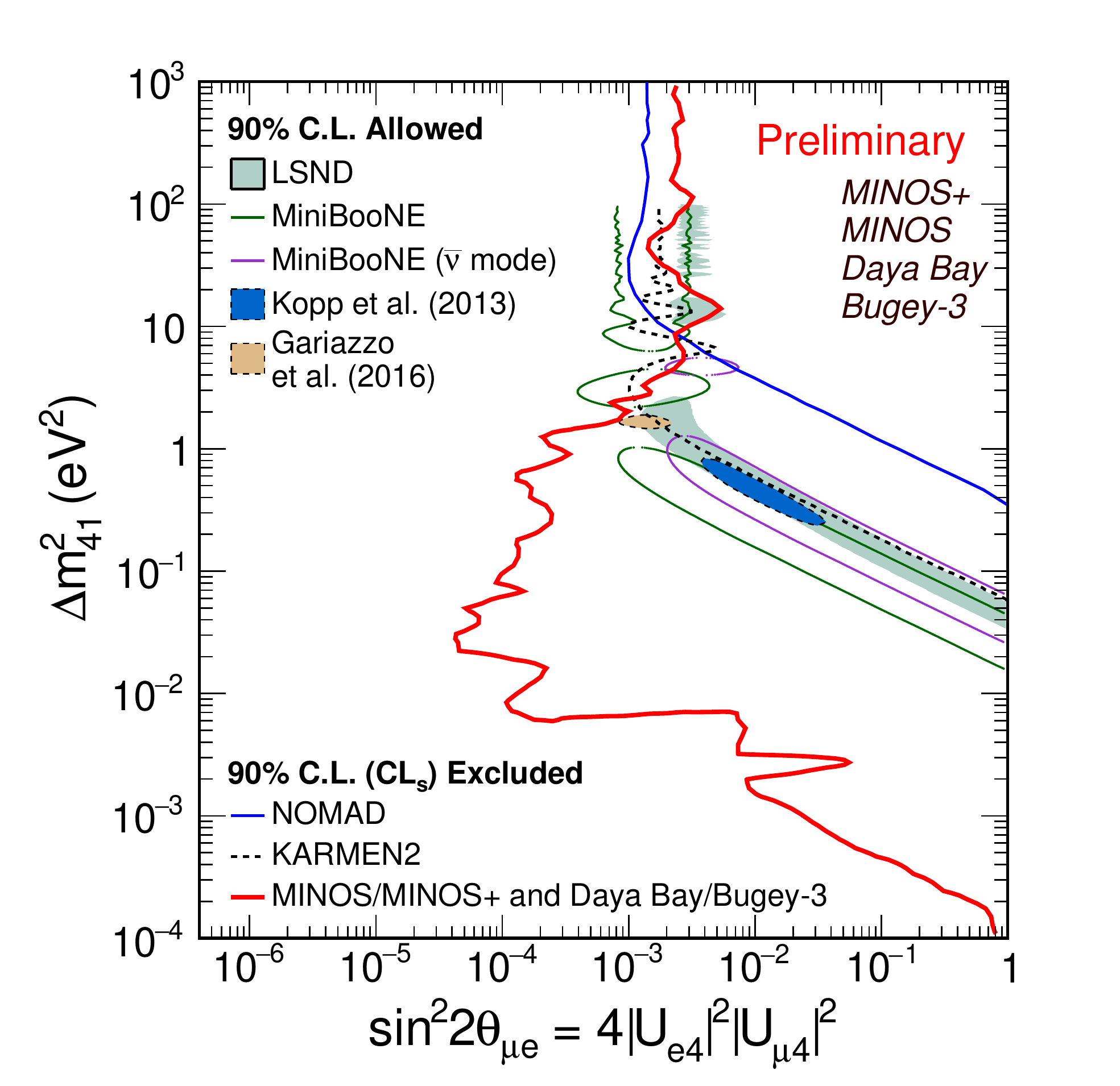}
\caption{The combined 90$\%$ C.L. limit on $\mathrm{sin^{2}2\theta_{\mu e}}$\cite{com}
. It excludes almost all of the 90$\%$
C.L. global allowed region.}
\label{fig:false-color}
\end{figure}

There are many experiments that have studied the $\nu_{e}$ appearance channel. These gave potential evidence for the existence of light sterile neutrinos. At the same time many experiments gave null results using $\nu_{e}$ appearance channel and $\nu_{\mu}$ disappearance channel. Figure $\ref{fig:false-color}$ shows the limit on mass and mixing from the combined analysis of MINOS+, Daya Bay and Bugey-3.

\section{\label{sec:level1}NO$\textnormal{v}$A Experiment}
The NOvA experiment is designed to study the neutrino oscillation properties. It consists of a ND and FD, which are functionally identical and 14.6 mrad off axis from the Fermilab NuMI beam. The ND is 1 km away from the the beam source. The ND and FD are separated by 809 km. The experiment is designed to operate in neutrino mode (using neutrino beam flux) and antineutrino mode (using antineutrino beam flux). The long base-line oscillation channels that studies in NOvA includes 1. $\nu_{e}$ appearance, 2. $\nu_{\mu}$ disappearance, 3. NC disappearance.
NOvA has the potential to measure the precise value of neutrino mixing angles, determine neutrino mass hierarchy and can investigate the CP violation in the lepton sector. It also has potential to provide constraints on exotic phenomenon such as sterile neutrino oscillation.

\section{\label{sec:level1}NC disappearance at the NO$\textnormal{v}$A FD}
The NC interactions are mediated by neutral $Z^{0}$ bosons. In this type of interaction, neutrino leaves the detector with reduced energy. NC interactions are distinct from the charged current (CC)
interactions, in the sense that they are devoid of a charged lepton. Only a fraction of the event energy is deposited as hadronic activity in the detector. 
\par
The basic method to search for active-sterile neutrino oscillation in NOvA FD is looking for the depletion of NC events compared to the predicted rate. Any difference between the ND data and simulation are accounted for by the FD prediction technique. Here ND data is extrapolated to the FD to make the prediction. Any further data Monte Carlo simulation difference is absorbed as systematic uncertainties.

\begin{figure}[tbp]
\centering
\includegraphics[width=\linewidth]{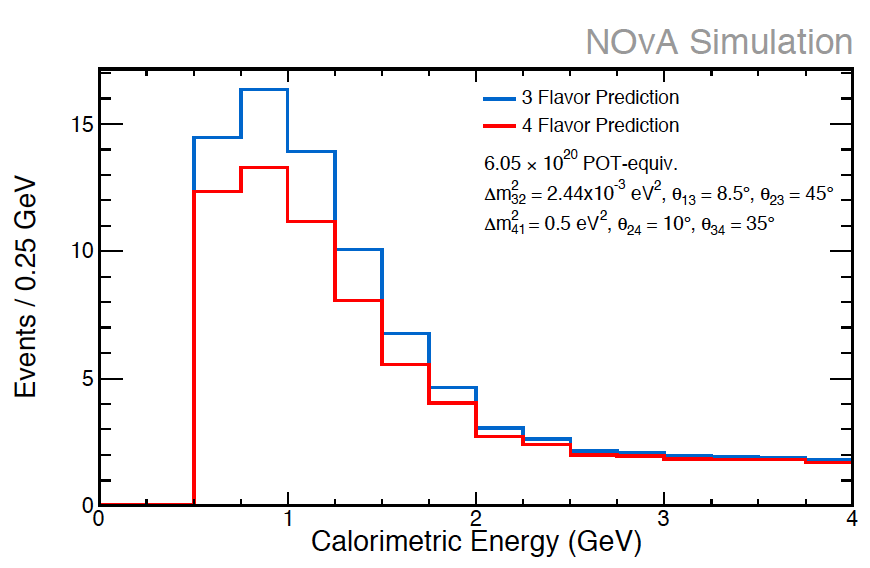}
\caption{NC depletion as a function of energy.}
\label{fig:nc_depletion}
\end{figure}
In this analysis we adopt a minimal ``3+1'' extension of three flavor neutrino model. Which adds $\Delta m^{\text{2}}_{\text{41}}$, $\theta_{14}$,$\theta_{24}$, $\theta_{34}$,$\delta_{14}$ and $\delta_{24}$ to the 3 flavor parameters. NOvA is sensitive to the measurement of the angles $\theta_{24}$,$\theta_{34}$ and $|U_{\tau4}|$ and $|U_{\mu4}|$. NC event depeletion as afunction of calorimetric energy is shown in $\ref{fig:nc_depletion}$.

\subsection{NC Event selection}
A computer vision based particle identifier, the Convolutional Visual Network (CVN)\cite{cvn} is used as primary selector for NC signals. Selected events should be well reconstructed, contained and in the fiducial volume of the detector. Along with CVN, a specific set of selection cuts and a boosted decision tree are employed for cosmic background rejection. Since the NOvA FD is at the ground level, an average of 148 kHz of cosmic rays are incident on the detector. So cosmic rays are a major background at the FD. After applying all selection criteria, in the FD(ND), we achieved a 50$\%$ (62$\%$) NC efficiency and (72 $\%$) 70 $\%$ NC signal purity.

\subsection{Extrapolating ND prediction to the FD}
The process of predicting the FD Calorimetric energy spectrum using ND energy spectrum can be summarized as: ND data is separated into CC and NC events and each component is extrapolated to FD by
 \begin{equation}
  ND^{Data}  \frac{FD^{MC}}{ND^{MC}} = FD^{Pred}
 \end{equation}
 Where $ND^{Data}$ is the ND data, $FD^{Pred}$ is the FD prediction and $\frac{FD^{MC}}{ND^{MC}}$ is the ratio of FD and ND Monte Carlo simulation.
Oscillation weights are applied to each FD predicted components. The extrapolated prediction is compared with the FD data.

\subsection{NC Disappearance first analysis results.}
We selected 95$\pm$9.7 NC candidates at FD, where as the predicted events are 83.5$\pm$0.8$ (\textit{stat.})^{\text{+10.9}}_{\text{-7.2}}$(syst.). The variable $R_{NC}$ is calculated as a model independent test for active to sterile mixing.
 \begin{equation}
  R_{NC} = \frac{F^{data} -\sum F^{pred}(bkg)}{F^{pred}(NC)}
 \end{equation}
 Here $F^{data}$ is the selected data at the FD. And $\sum F^{pred}(bkg)$ is the sum of the predictions of backgrounds and $F^{pred}(NC)$ is the NC signal prediction at the FD. The predicted quantities are calculated based on the 3-flavor oscillation model. If NC disappearance occurs $R_{NC}$ $<$ 1. But we see $R_{NC}$ =1.19 $\pm$ 0.16(stat.) + 10 (syst.), which is a 1.03 sigma excess of NC events consistent with standard 3-flavor oscillations. The measurement is consistent with the 3-flavor oscillation model.

\begin{figure}[tbp]
\centering
\includegraphics[width=\linewidth]{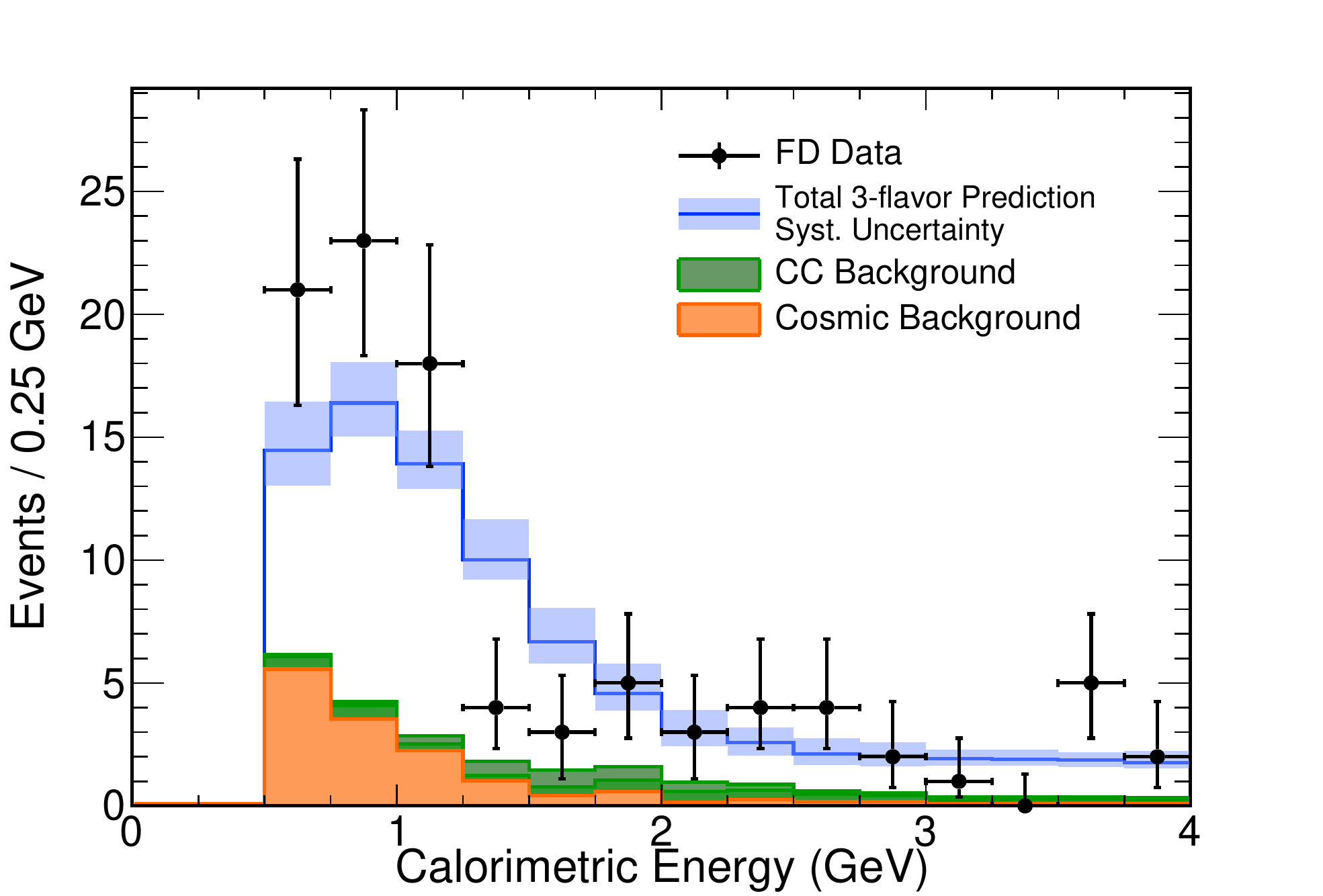}
\caption{The Calorimetric Energy comparison of  data and MC predicted events at the FD for 6.05x$10^{20}$ POT equivalent. }
\label{fig:fd_events}
\end{figure}

Using 3+1 model, we fit the data for $\theta_{24}$ and $\theta_{34}$ using the same oscillation parameter values and uncertainties as for the 3-flavor neutrino oscillation prediction. This analysis has some sensitivity to  $\delta_{24}$ . So this is profiled over. Then the angle is determined by minimizing the $\chi^{2}$. The 90$\%$ C.L limit values for $\theta_{24}$ $<20.8$ and $\theta_{34}$ $<31.2$ are obtained. This values can be converted into respective matrix elements as $|{U_{\mu4}}|^{2}$ $<$ 0.126 and $|{U_{\tau4}}|^{2}$ $<$ 0.268 at the 90 $\%$ C.L.

\begin{figure}[tbp]
\centering
\includegraphics[width=\linewidth]{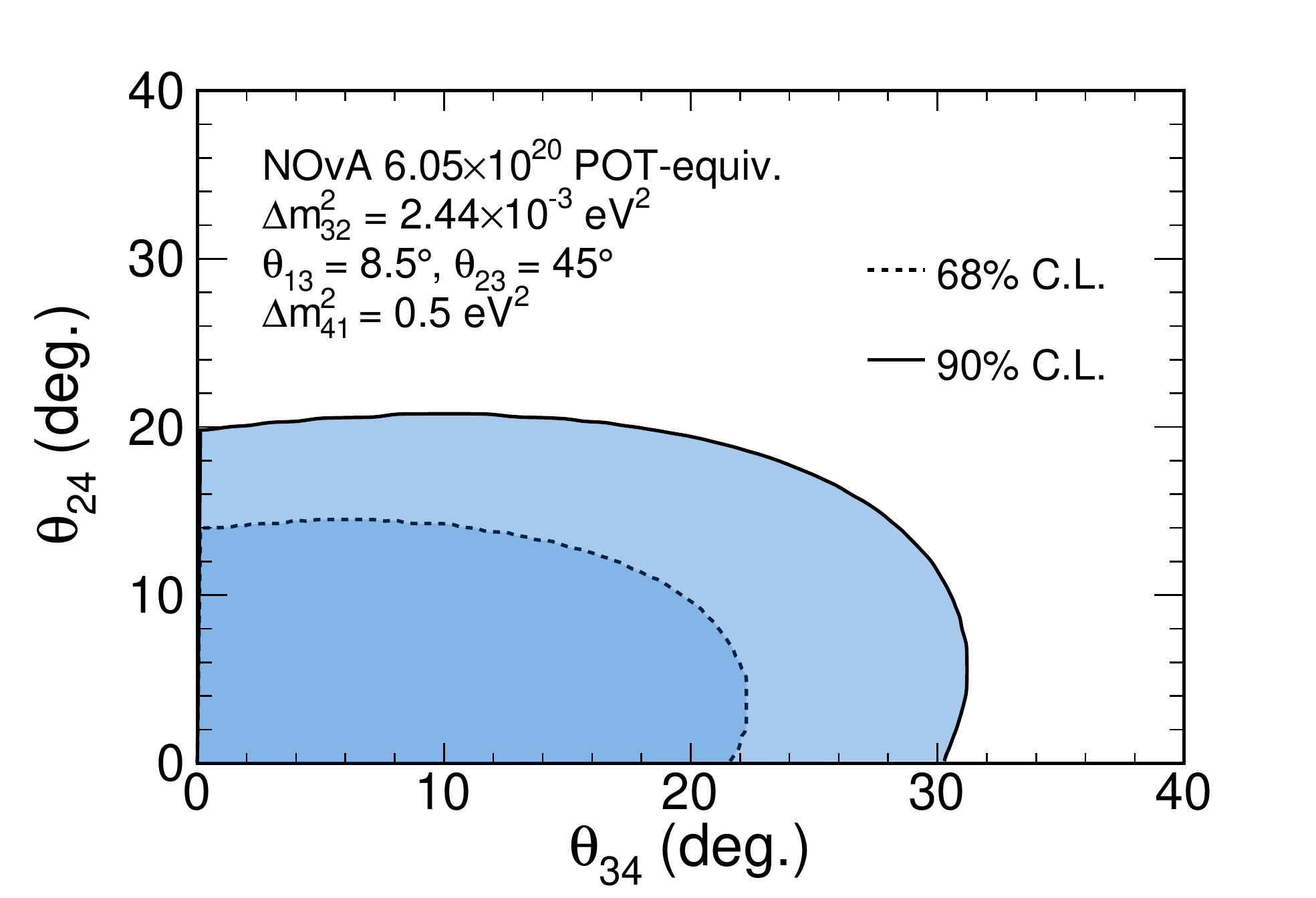}
\includegraphics[width=\linewidth]{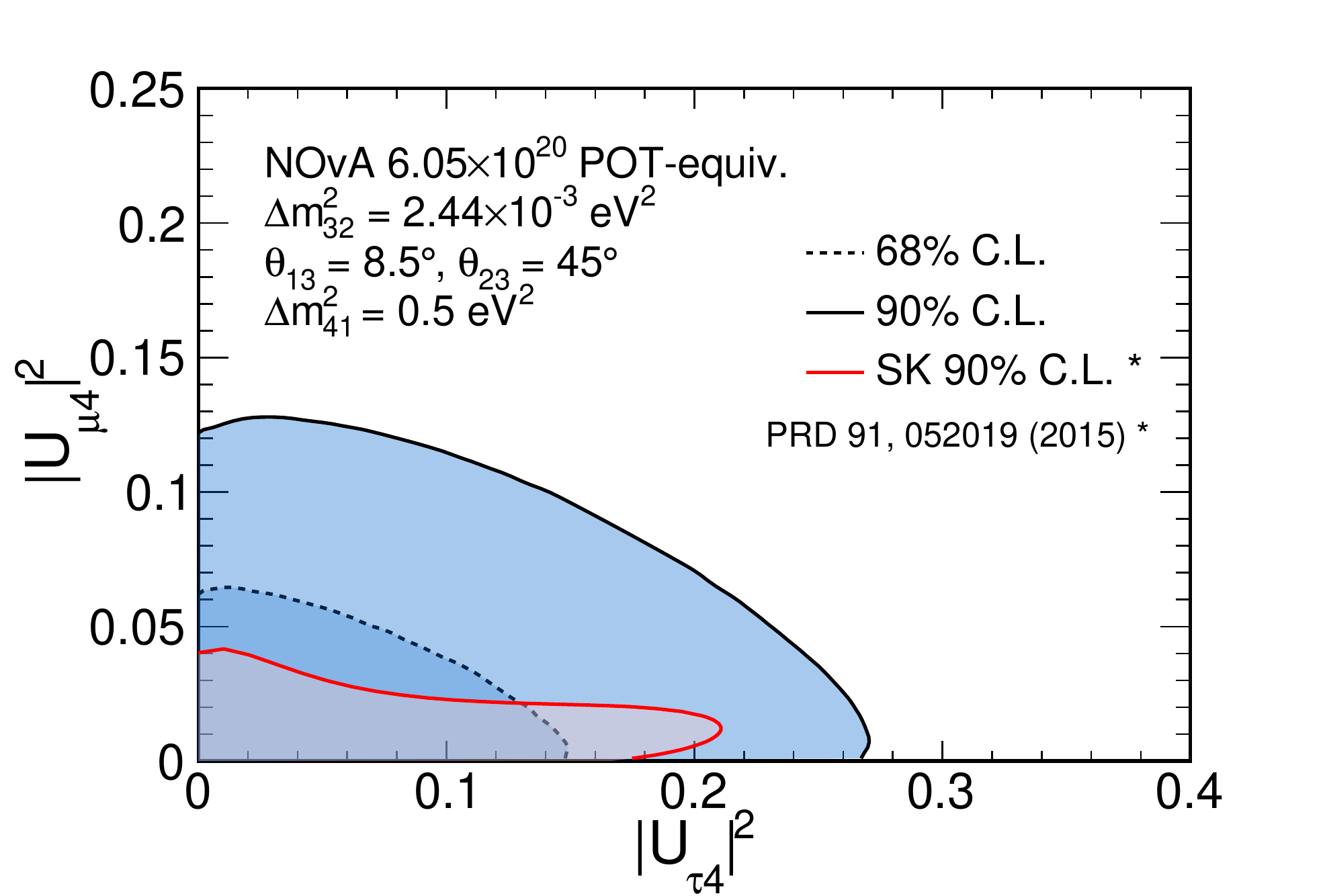}
\caption{Top: The 68$\%$ (dashed) and 90$\%$ (solid) Feldman-Cousins non-excluded regions for mixing angles $\theta_{24}$ and $\theta_{34}$.  Bottom: The 68 $\%$ (dashed) and 90 $\%$ (solid) Feldman-Cousins non-excluded regions for  $|{U_{\mu4}}|^{2}$ and $|{U_{\tau4}}|^{2}$  assuming $\mathrm{cos^{2}\theta_{14}}$ =1 in both cases.}
\label{fig:fd_}
\end{figure}

\begin{table} 
\begin{tabular}{l | c | c | c |c}
\hline
 Parm. & NOvA & SuperK& MINOS & IceCube \\
\hline \textbf{$\theta_{34}$} & $31.2^{0}$ & $25.1^{0}$ & $26.6^{0}$ &  -\\ 
\hline \textbf{$\theta_{24}$} & $20.8^{0}$ & $11.7^{0}$ & $7.3^{0}$ & $4.1^{0}$\\ 
\hline \textbf{$|U_{\mu4}|^{2}$} & 0.126 & 0.041 & 0.016 & 0.005 \\
\hline \textbf{$|U_{\tau4}|^{2}$} & 0.268 & 0.18& 0.20 & -\\
\hline
\end{tabular}
\caption{Comparing the 90 $\%$ C.L limit of NOvA first NC disappearance analysis result with SuperK\cite{superk}, MINOS\cite{minos} and IceCube\cite{ice}. The limits are shown for $\Delta m_{41}^{2}$ = 0.5 e$V^{2}$ for all experiments }
\label{table:t1}
\end{table}
\section{Improvements for the future Analyses}
\subsection{Shape fit}
The 2016 analysis was a counting experiment, which compares the FD NC event rate with predicted rate and fitted for the total event count. Improved modeling of the detector and the neutrino interaction cross-section will enable for a shape fit. These improvements include, improved detector modeling, more accurate threshold modeling from data, Cherenkov light modeling and
the improved cross-section modeling of the NC events.
In contrast to the rate fit, the shape information of the FD spectrum is input to the shape fit.
\subsection{Visible energy correction}

The signature of the NC event in the detector is hadronic energy deposition. The remaining energy is carried away by the out-going neutrino. The reconstructed NC event energy is corrected by the single parameter correction method. 

\subsection{Extending the mass square splitting range.}
In the 2016 analysis, $\Delta m_{41}^{2}$ range was restricted to between 0.05 $eV^{2}$ and 0.5 $eV^{2}$. In this range the analysis is not sensitive to the oscillations which affect the rate in the ND. 
This was because oscillations in the ND are not considered. The addition of ND oscillation will enable the analysis to expand the $\Delta m_{41}^{2}$ range.

\begin{figure}[tbp]
\centering
\includegraphics[width=\linewidth]{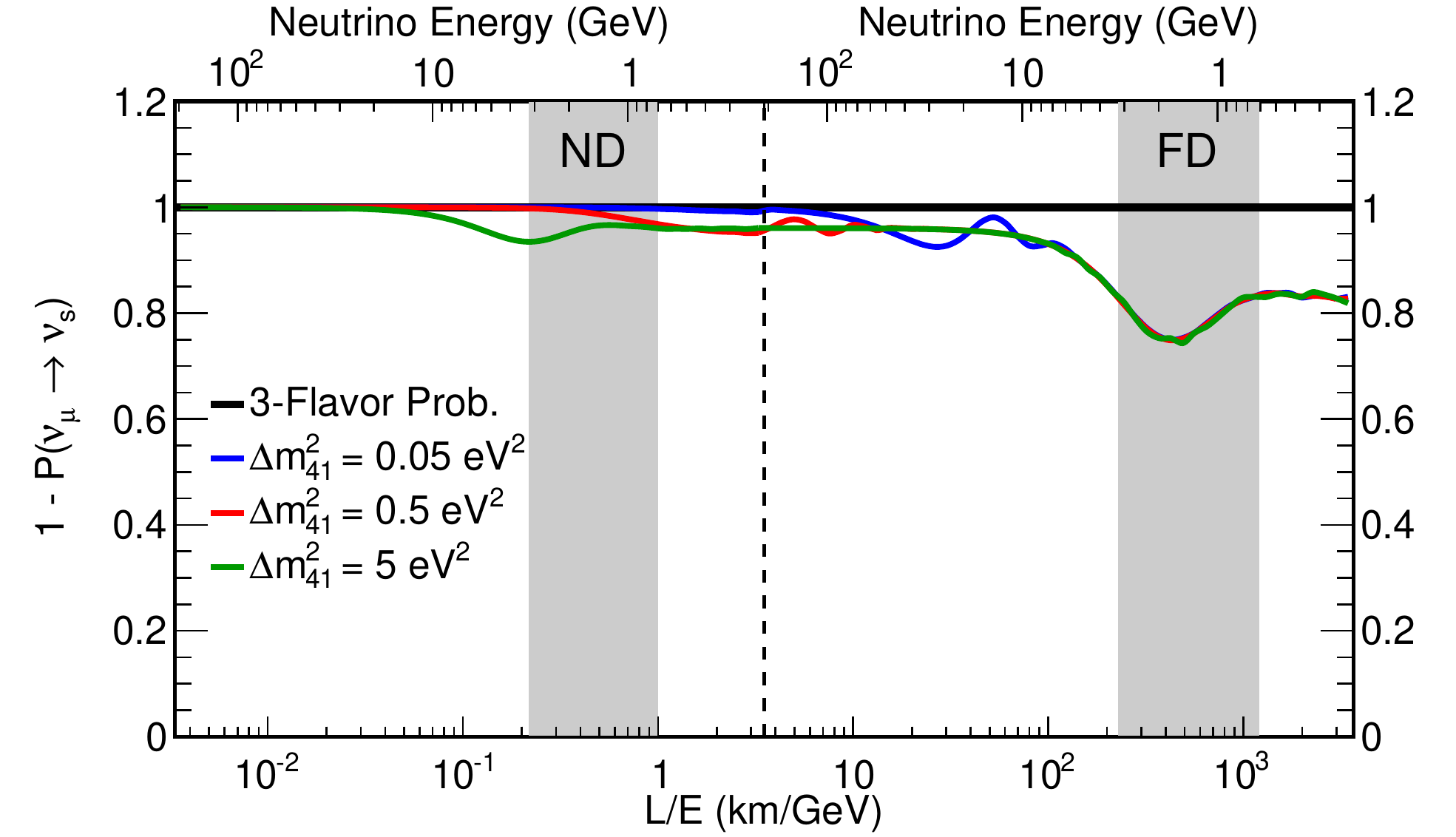}
\caption{The plot shows the change of the oscillation probability with different $\Delta m_{41}^{2}$}
\label{fig:osc}
\end{figure}
\section{conclusion}
For the NC disappearance 2016 analysis, 6.69x$10^{20}$ POT was collected at the FD. The 2017 analysis will add 50$\%$ more data. 
The analysis improvements and increased data will enable NOvA to set compititive limit on the best-fit values of the $\theta_{34}$ and $|U_{\tau4}|^{2}$ within a few years. An antineutrino run will also open new analysis possibilities.  

\section{Acknowledgements}
The NOvA collaboration uses the resources of the Fermi National Accelerator Labo-
ratory (Fermilab), a U.S. Department of Energy, Office of Science, HEP User Facility.
Fermilab is managed by Fermi Research Alliance, LLC (FRA), acting under Contract
No.  DE-AC02-07CH11359.  This research was supported by the U.S. Department
of Energy; the U.S. National Science Foundation; the Department of Science and
Technology, India; the European Research Council; the MSMT CR, GA UK, Czech
Republic; the RAS, RMES, and RFBR, Russia; CNPq and FAPEG, Brazil; and the
State and University of Minnesota. We are grateful for the contributions of the staff
at Fermilab and the NOvA Far Detector Laboratory.

\end{document}